# Truss Decomposition in Massive Networks

Jia Wang
Department of Computer Science and Engineering
The Chinese University of Hong Kong
jwang@cse.cuhk.edu.hk

James Cheng
School of Computer Engineering
Nanyang Technological University, Singapore
j.cheng@acm.org

## ABSTRACT

The $k$-truss is a type of cohesive subgraphs proposed recently for the study of networks. While the problem of computing most cohesive subgraphs is NP-hard, there exists a polynomial time algorithm for computing $k$-truss. Compared with $k$-core which is also efficient to compute, $k$-truss represents the "core" of a $k$-core that keeps the key information of, while filtering out less important information from, the $k$-core. However, existing algorithms for computing $k$-truss are inefficient for handling today's massive networks. We first improve the existing in-memory algorithm for computing $k$-truss in networks of moderate size. Then, we propose two I/O-efficient algorithms to handle massive networks that cannot fit in main memory. Our experiments on real datasets verify the efficiency of our algorithms and the value of $k$-truss.

## 1. INTRODUCTION

Given a graph $G$, the **$k$-truss** of $G$ is *the largest subgraph of $G$ in which every edge is contained in at least $(k-2)$ triangles within the subgraph* [15, 16]. The problem of **truss decomposition** in $G$ is to find the (non-empty) $k$-trusses of $G$ for all $k$.

The $k$-truss is a type of *cohesive subgraphs* (or *cohesive groups*) of a network [18]. In the analysis of a massive network, it is often more fruitful, and more feasible, due to the large size of the network, to focus on smaller but more important areas of the network, i.e., subgraphs that reflect, and/or can be used to study, important properties of the network such as connectivity, robustness, self-similarity, centrality, etc. Thus, network analysts have attempted to identify various cohesive subgraphs for more efficient and effective analysis of a network.

In the literature, many notions of cohesive subgraphs were proposed. The basic one are the *cliques* (i.e., a subset of vertices that forms a complete subgraph) [21] and *maximal cliques* [7]. However, the definition of clique is often too rigid and thus other more relaxed forms of cohesive subgraphs were proposed. The *n-clique* [22] relaxes the distance between any two vertices in a clique from 1 to $n$. The *k-plex* [29] relaxes the degree of each vertex within a clique of $c$ vertices from $(c-1)$ to $(c-k)$. The *n-clan* [24] is the same as the $n$-clique except for imposing a constraint on the diameter, while the *n-club* [24] removes the $n$-clique requirement from the $n$-clan. The *quasi-clique* can be either a relaxation on the density [1] or the degree [23, 26]. However, the computation of all the above cohesive subgraphs is NP-hard.

All the above-mentioned cohesive subgraphs are relatively small sub-structures in a graph. They may be scattered all over the graph, and some of them may overlap largely with each other, while others are disconnected from each other. For example, two cliques may share all but one vertex with each other, or may be totally isolated from each other.

On the contrary, $k$-trusses are hierarchical subgraphs that represent the cores of a network at different levels of granularity. In this sense, $k$-truss is more similar to $k$-core [28], which is the largest subgraph of a graph in which every vertex has degree at least $k$ within the subgraph. However, the $k$-core was described by Seidman [28] as a "seedbed" within which cohesive subgraphs may precipitate (e.g. a $k$-truss is a $(k-1)$-core but not vice versa) [15]. Thus, though containing other cohesive subgraphs such as cliques and $k$-truss, the $k$-core may also contain a lot more that can be further filtered out for more effective and efficient network analysis.

Conceptually, the definition of $k$-truss is also more rigorous than that of $k$-core since $k$-truss is defined based on *triangles*, which are known as fundamental building blocks of a network [32, 33, 25, 6]. In a social network, a triangle implies a strong tie among three friends, or two friends having a common friend. Thus, by enforcing all edges to be contained in at least $(k-2)$ triangles, the $k$-truss strengthens every (edge) connection in it by at least $(k-2)$ strong ties. Intuitively, we may consider this as in a social network, the more common friends two people have, the stronger their connection it implies. On the contrary, in a $k$-core we only have simple edge connection (i.e., degree).

We give an example to illustrate the difference between $k$-truss and $k$-core as follows.

*Example 1.* Figure 1(a) shows a graph, $G$, that displays the "seek-advice-from" relationship among the managers of a high-tech manufacturing firm in the west coast of the U.S. [19, 32, 15]. Figure 1(b) and 1(c) show the 3-core and 4-truss of $G$, which are also given as an example in [15]. Note that no 4-core or 5-truss exist for $G$, i.e., there is no subgraph of $G$ that has all vertices with degree at least 4 or all edges contained in at least $(5-2) = 3$ triangles within the subgraph.

The figures show that the 3-core is not much different from the original graph $G$, while the 4-truss is significantly different from $G$. The clustering coefficient [33] of $G$, of the 3-core, and of the 4-truss is calculated to be 0.51, 0.65, and 0.80, respectively, indicating that the 4-truss tends to form clusters or communities at a degree much higher than $G$ and the 3-core.

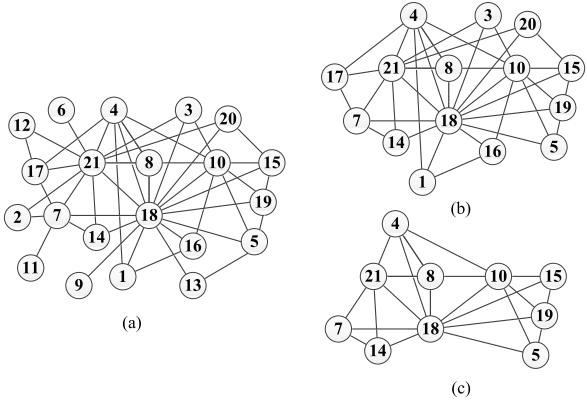

**Figure 1: (a)** A manager-relationship graph, $G$; **(b)** the 3-core of $G$ (no 4-core exists); **(c)** the 4-truss of $G$ (no 5-truss exists)

The 4-truss also satisfies the requirement of a 3-core by definition (not vice versa); however, the 4-truss further filters out those vertices with lower local clustering coefficient in the 3-core, i.e., those vertices that do not tend to form clusters or tightly-knit community structures with others.

From another angle, we may also see that the 4-truss contains all the cliques with more than 3 vertices, namely {4,8,10,18}, {4,8,18,21}, {5,10,18,19}, {7,14,18,21}, and {10,15,18,19}, but it filters out all the other less cohesive sub-structures that exist in the 3-core.

This example also shows that, as the "*core*" part of the $k$-core, the $k$-truss has a smaller size and a clearer display of the essential part of a network. Thus, $k$-truss can be more suitable than $k$-core in applications such as visualization and fingerprinting of large-scale networks [3], interpretation of cooperative processes in complex networks [8], and analysis of network connectivity [4], etc. □

On the computational aspect, the non-hierarchical cohesive subgraphs are mainly evolved from the cliques and therefore expensive to compute. On the contrary, computing the hierarchical cohesive subgraphs, $k$-core and $k$-truss, has a polynomial time complexity.

Cohen [15] proposed an algorithm for truss decomposition, which requires random access to vertices/edges and hence the entire input graph to be resident in main memory. However, real world networks have grown drastically in recent years. It is unrealistic to assume that these graphs/networks can always fit in main memory.

Recently, Cohen [16] also proposed a parallel algorithm based on the MapReduce framework which does not require keeping the entire input graph in any single machine. To compute the $k$-truss, the algorithm iteratively invokes a MapReduce procedure for triangle listing whenever there are some edges that are contained in less than $(k-2)$ triangles. The iterative counting of triangles enforced in the definition of $k$-truss requires many iterations of a main procedure that makes parallelization of the entire process difficult.

In this paper, we first propose an efficient in-memory algorithm for truss decomposition that has the same worst-case complexity as the *lower-bound* complexity of in-memory triangle listing [30]. We also show that our algorithm is significantly more efficient than the existing algorithms for truss decomposition [15, 16] in small networks. However, for processing massive networks that cannot fit in main memory, the existing algorithms become impractical due to huge I/O cost.

We develop two I/O-efficient algorithms for truss decomposition in massive networks that cannot fit in memory. The first one is a bottom-up approach that employs an effective pruning strategy by removing a large portion of edges before the computation of each $k$-truss, thus significantly reducing both disk I/O cost and search space during truss decomposition. The second one takes a top-down approach, which is tailor-made for applications that prefer the $k$-trusses of larger values of $k$, as they often represent the heart or backbone of a network.

We evaluate our algorithms on a range of real world networks. For networks of moderate sizes that can reside in memory, the results show that our in-memory algorithm significantly improves the existing in-memory algorithm [15]. For larger networks that cannot fit in memory, the results show that our I/O-efficient method is much more efficient than the existing MapReduce algorithm [16].

**Organization.** Section 2 formally defines the problem and gives the basic notations. Section 3 describes the in-memory algorithms and identifies their limitations. Section 4 gives an overview of the two I/O-efficient algorithms and highlights the challenges. Sections 5 and 6 discuss in details the bottom-up and top-down approaches. Section 7 reports the experimental results. Section 8 discusses the related work and Section 9 concludes the paper.

## 2. PROBLEM DEFINITION

We consider undirected, unweighted simple graphs. Given a graph $G$, we denote by $V_G$ and $E_G$ the vertex set and edge set of $G$, respectively. We define $n = |V_G|$ as the number of vertices, and $m = |E_G|$ as the number of edges of $G$. We define the size of $G$, denoted by $|G|$, as $|G| = m + n$. We use $nb(v)$ to denote the set of *neighbors* of a vertex $v$, that is, $nb(v) = \{u : (u,v) \in E_G\}$. We define the *degree* of $v$ in $G$ as $deg(v) = |nb(v)|$.

Unless otherwise specified, we assume that a graph is stored in its adjacency list representation (whether in memory or on disk). Each vertex in the graph is assigned a unique ID. In the adjacency list representation, vertices are sorted in ascending order of their IDs. Given any two vertices $u$ and $v$, we use $u < v$ or $v > u$ to denote that $u$ is ordered before $v$.

A *triangle* in $G$ is a cycle of length 3. Let $u, v, w \in V_G$ be the three vertices on the cycle, we denote this triangle by $\triangle_{uvw}$. In addition, we denote the set of triangles of $G$ by $\triangle_G$. On the basis of triangles, we define the *support* of an edge [15] as follows.

*Definition* 1 (SUPPORT). *The* support *of an edge* $e = (u,v) \in E_G$ *in* $G$, *denoted by* $sup(e, G)$, *is defined as* $|\{\triangle_{uvw} : \triangle_{uvw} \in \triangle_G\}|$. *When* $G$ *is obvious from content, we replace* $sup(e, G)$ *by* $sup(e)$.

The support of an edge $e$ in $G$ is simply the number of triangles in $G$ that contain $e$. Now we define the notion of $k$-truss [15, 16].

*Definition* 2 ($k$-TRUSS). *The* $k$-truss *of* $G$, *where* $k \geq 2$, *denoted by* $T_k$, *is defined as the largest subgraph of* $G$, *such that* $\forall e \in E_{T_k}, sup(e, T_k) \geq (k-2)$.

By definition, the 2-truss is simply $G$ itself.

We define the *truss number*, or *trussness*, of an edge $e$ in $G$, denoted by $\phi(e)$, as $\max\{k : e \in E_{T_k}\}$. It follows that given $\phi(e) = k$, we have $e \in E_{T_k}$ but $e \notin E_{T_{k+1}}$. We use $k_{max}$ to denote the maximum truss number of any edge in $G$. From the truss number comes our definition of the $k$-class.

*Definition* 3 ($k$-CLASS). *The* $k$-class *of* $G$, *denoted by* $\Phi_k$, *is defined as* $\{e : e \in E_G, \phi(e) = k\}$.

**Problem definition.** The problem of **truss decomposition** studied in this paper is defined as follows. *Given a graph $G$, compute*

| Notation | Description |
|---|---|
| $G = (V_G, E_G)$ | A undirected, unweighted simple graph $G$ |
| $n; m$ | The number of vertices/edges in $G$ |
| $|G|$ | The size of $G$, $|G| = m + n$ |
| $nb(v)$ | The set of neighbors of $v \in V_G$ |
| $deg(v)$ | The degree of vertex $v \in V_G$ |
| $\triangle_{uvw}$ | A triangle formed by $u$, $v$ and $w$ |
| $sup(e)$ | The support of $e$ in $G$ |
| $sup(e, H)$ | The support of $e$ in a subgraph $H$ of $G$ |
| $T_k$ | The $k$-truss of $G$ |
| $\phi(e)$ | The truss number of $e$ in $G$ |
| $\phi(e, H)$ | The truss number of $e$ in a subgraph $H$ of $G$ |
| $\Phi_k$ | The $k$-class of $G$, $\Phi_k = \{e : e \in E_G, \phi(e) = k\}$ |
| $\varphi(e)$ | The estimated lower bound of $\phi(e)$ for $e \in E_G$ |
| $\psi(e)$ | The estimated upper bound of $\phi(e)$ for $e \in E_G$ |
| $M$ | The size of available main memory |
| $B$ | The disk block size |
| $scan(N)$ | $\Theta(N/B)$ |

Table 1: Frequently Used Notations

the $k$-truss of $G$ for all $2 \leq k \leq k_{max}$. Equivalently, the $k$-truss can be obtained from the set of edges $E_{T_k} = \bigcup_{j \geq k} \Phi_j$, i.e., the union of all edges with truss number at least $k$. When the input graph $G$ cannot fit in memory, we propose I/O-efficient algorithms to compute the $k$-classes.

The following example illustrates the concept of $k$-truss.

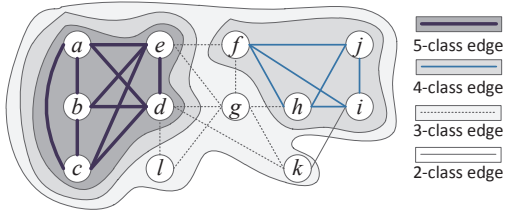

Figure 2: A graph $G$ and the $k$-classes of $G$ ($2 \leq k \leq 5$)

*Example 2.* In the graph $G$ in Figure 2, the different types of edges indicate the different $k$-classes, where $2 \leq k \leq 5$. In $G$, the 2-class $\Phi_2$ has a single edge $(i, k)$, i.e., $\Phi_2 = \{(i, k)\}$, since $(i, k)$ is the only edge in $G$ with support 0. The 3-class $\Phi_3$ consists of 9 edges, given by $\Phi_3 = \{(d, g), (d, k), (d, l), (e, f), (e, g), (f, g), (g, h), (g, k), (g, l)\}$. The 4-class $\Phi_4$ contains 6 edges, given by $\Phi_4 = \{(f, h), (f, i), (f, j), (h, i), (h, j), (i, j)\}$. The 5-class $\Phi_5$ consists of 10 edges, given by $\Phi_5 = \{(a, b), (a, c), (a, d), (a, e), (b, c), (b, d), (b, e), (c, d), (c, e), (d, e)\}$. We have $k_{max} = 5$.

From the $k$-classes we obtain the $k$-trusses as follows. The 2-truss $T_2$ is simply $G$ itself. The 3-truss $T_3$ is the subgraph of $G$ formed by the edge set $(\Phi_3 \cup \Phi_4 \cup \Phi_5)$. The 4-truss $T_4$ is the subgraph formed by $(\Phi_4 \bigcup \Phi_5)$, and the 5-truss $T_5$ is the subgraph formed by $\Phi_5$. We can verify that for $2 \leq k \leq 5$, each edge of $T_k$ is contained in at least $k-2$ triangles in $T_k$. The $k$-trusses display the hierarchical structures of $G$ at different levels of granularity, as depicted by the shading of different gray scales in Figure 2. □

Table 1 lists the notations that are frequently used in the paper. When analyzing I/O complexity of our algorithms, we adopt the I/O model in [2]: $M$ is the main memory size and $B$ is the disk block size, where $1 \ll B \leq M/2$. Data is read/written in blocks from/to disk. Thus, reading/writing a piece of data of size $N$ from/to disk requires $(N/B)$ I/Os. We also define $scan(N) = \Theta(\frac{N}{B})$, where $N$ is the amount of data being read or written from/to disk.

**Algorithm 1** *Truss Decomposition*

**Input**: $G = (V_G, E_G)$
**Output**: the $k$-truss for $3 \leq k \leq k_{max}$

1. $k \leftarrow 3$;
2. **for** each $e = (u, v) \in E_G$ **do**
3. $\quad sup(e) = |nb(u) \cap nb(v)|$;
4. **while**($\exists e = (u, v)$ such that $sup(e) < (k-2)$)
5. $\quad W \leftarrow (nb(u) \cap nb(v))$;
6. $\quad$ **for** each $e' = (u, w)$ or $e' = (v, w)$, where $w \in W$, **do**
7. $\quad\quad sup(e') \leftarrow (sup(e') - 1)$;
8. $\quad$ remove $e$ from $G$;
9. output $G$ as the $k$-truss;
10. **if**(*not* all edges in $G$ are removed)
11. $\quad k \leftarrow (k + 1)$;
12. $\quad$ **goto** Step 4;

## 3. IN-MEMORY TRUSS DECOMPOSITION

We first describe an existing algorithm for truss decomposition. Then, we propose an improved algorithm and prove that its time complexity is bounded by that required for triangle listing. Finally, we identify the limitation of the in-memory algorithms.

### 3.1 The Existing In-Memory Algorithm

We outline the algorithm of [15] for truss decomposition in Algorithm 1. The algorithm starts with an initialization by computing the support of every edge in $G$ (Steps 2-3). The intersection of $nb(u)$ and $nb(v)$ for each edge $e = (u, v)$ returns the set of vertices that form triangles with $u$ and $v$, and thus the cardinality of the intersection gives the support of $e$.

After the initialization, for each $k$ starting from $k = 3$, the algorithm removes every edge $e = (u, v)$ with support less than $(k-2)$, since $e$ cannot be in the $k$-truss by definition (Steps 4-8). Removing $e$, however, may also invalidate all triangles consisting of $e$, i.e., $\forall \triangle_{uvw}$, where $w \in W = (nb(u) \cap nb(v))$, $\triangle_{uvw}$ is no longer a valid triangle after the removal of $e = (u, v)$. Thus, we also decrement the support of the other two edges $(u, w)$ and $(v, w)$ for each $\triangle_{uvw}$, where $w \in W$. This process is repeated iteratively until all the remaining edges in $G$ have support at least $(k - 2)$, which is the $k$-truss.

If there are still some edges in $G$ not yet removed, we continue with the next $k$ by repeating the above process, i.e., Steps 4-9.

In Step 4 of Algorithm 1, we need to find edges with support less than $(k - 2)$. This step can be efficiently processed by using a queue. Whenever we compute (during initialization) or update (upon the removal of an edge) the support of an edge, we push the edge into the queue if its support becomes less than $(k - 2)$ or update its position in the queue if the edge is already in the queue.

In Step 8 of Algorithm 1, we remove edges from $G$. Explicitly deleting edges from $G$ during every step of the process can be expensive since it involves updating the adjacency lists of $u$ and $v$ for each edge $e = (u, v)$, which requires time linear in $(deg(u) + deg(v))$. Thus, an implicit approach by simply marking that $e$ has been deleted in $nb(u)$ and $nb(v)$ is more efficient.

**Complexity.** Algorithm 1 requires $O(m + n)$ memory space to keep the input graph as well as the support of all edges in memory. The initialization (Steps 2-3) can be made faster using the in-memory triangle counting algorithm [27, 20]. However, Step 5 still requires $O(deg(u) + deg(v))$ time for each edge $e = (u, v)$ processed, thus giving a total of $O(\sum_{(u,v) \in E_G}(deg(u) + deg(v)))$ $= O(\sum_{v \in V_G}(deg(v))^2)$ time. This can be expensive for large graphs with vertices of high degree.

**Algorithm 2** *Improved Truss Decomposition*

**Input**: $G = (V_G, E_G)$
**Output**: the $k$-class, $\Phi_k$, for $2 \leq k \leq k_{max}$

1. $k \leftarrow 2$, $\Phi_k \leftarrow \emptyset$;
2. compute $sup(e)$ for each edge $e \in E_G$;
3. sort all the edges in ascending order of their support;
4. **while**($\exists e$ such that $sup(e) \leq (k-2)$)
5.    let $e = (u, v)$ be the edge with the lowest support;
6.    assume, w.l.o.g., $deg(u) \leq deg(v)$;
7.    **for** each $w \in nb(u)$ **do**
8.      **if**$((v, w) \in E_G)$
9.        $sup((u, w)) \leftarrow (sup((u, w)) - 1)$,
    $sup((v, w)) \leftarrow (sup((v, w)) - 1)$;
10.      reorder $(u, w)$ and $(v, w)$ according to their new support;
11.    $\Phi_k \leftarrow (\Phi_k \cup \{e\})$;
12.    remove $e$ from $G$;
13. **if**(*not* all edges in $G$ are removed)
14.    $k \leftarrow (k+1)$;
15.    **goto** Step 4;
16. **return** $\Phi_j$, for $2 \leq j \leq k$;

## 3.2 An Improved Algorithm

We now present an improved algorithm for truss decomposition, as shown in Algorithm 2.

The algorithm also starts from an initialization that computes the support of each edge in $G$. We then sort all the edges in ascending order of their support. The computation of the support of the edges can be done in $O(m^{1.5})$ time by the in-memory triangle counting algorithm [27, 20]. The sorting can be done in $O(m)$ time with $O(m)$ space using bin sort. The sorted edges are then kept in an array $\mathcal{A}$, in a way similar to the sorted degree array in [5].

After the initialization, for each $k$ starting from $k = 2$, the algorithm iteratively removes a lowest support edge $e = (u, v)$, i.e., the first edge in the sorted edge array $\mathcal{A}$, while the support of $e$ is not greater than $k - 2$. The removed edge $e$ is added to $\Phi_k$ since $sup(e) \leq (k-2)$ and thus $e$ cannot be in $\Phi_{k+1}$. Upon the removal of $e$, we also decrement the support of all other edges that form a triangle with $e$, and update their new positions in the sorted edge array $\mathcal{A}$. The membership test of whether $(v, w) \in E_G$ at Step 8 can be done efficiently by keeping $E_G$ in a hashtable. The location of each edge in $\mathcal{A}$ is stored in the hashtable. Each update can be done in $\mathcal{A}$ in constant time in a way similar to that in the sorted degree array in [5]. We do not explicitly remove $e$ from $G$, but simply move a pointer one position forward in $\mathcal{A}$ to point to the next edge with the lowest support (i.e., all edges to the left of the pointer have been removed). This process continues until all edges with support less than or equal to $(k-2)$ are removed.

In this way, we compute each $k$-class until all edges in $G$ are removed.

Algorithm 2 is similar to Algorithm 1 but there is one subtle difference between Steps 5-6 of Algorithm 1 and Steps 6-8 of Algorithm 2 in the search for $\triangle_{uvw}$. This difference significantly reduces the time complexity of Algorithm 2 as shown in Theorem 1. We also show by experiments that Algorithm 2 is indeed significantly faster than Algorithm 1.

The correctness of Algorithm 2 is apparent since the algorithm essentially computes the $k$-truss by its definition. We show the complexity of this algorithm as follows.

THEOREM 1. *Algorithm 2 computes the $k$-truss, for all $k \geq 3$, in $O(m^{1.5})$ time using $O(m + n)$ space.*

PROOF. The support computation at Step 2 uses $O(m^{1.5})$ time and $O(m + n)$ space [27, 20], while Step 3 uses $O(m)$ time with $O(m)$ space using bin sort.

In the main iterative loop (i.e., Steps 4-12), Step 8 can be done in expected constant time by hashing and all other individual operations at other steps (except Step 7) can done in constant time (see the details in the algorithm description above). Thus, the total time depends on how many times these individual operations are executed. Steps 11-12 are obviously executed $O(m)$ times. Therefore, we only need to analyze how many times the individual operations at Steps 8-10 are executed.

For each edge $e = (u, v)$ removed, where $deg(u) \leq deg(v)$, as indicated by Step 6, the operations at Steps 8-10 are executed for at most $deg(u)$ times. Let $nb_\geq(u)$ be $\{v : v \in nb(u), deg(v) \geq deg(u)\}$. Thus, the loop at Steps 7-10 is executed for $|nb_\geq(u)|$ times for the vertex $u$, and the operations at Steps 8-10 are executed for at most $(deg(u) \cdot |nb_\geq(u)|)$ times.

We prove that for any $u \in V_G$, $|nb_\geq(u)| \leq 2\sqrt{m}$. If $deg(u) \leq \sqrt{m}$, then trivially $|nb_\geq(u)| \leq deg(u) \leq 2\sqrt{m}$. If $deg(u) > \sqrt{m}$, we prove by contradiction. Suppose $|nb_\geq(u)| > 2\sqrt{m}$. Then, $\sum_{v \in nb_\geq(u)} deg(v) \geq (|nb_\geq(u)| \cdot deg(u)) > (2\sqrt{m} \cdot \sqrt{m}) = 2m$, contradicting the fact that $\sum_{v \in V_G} deg(v) = 2m$. Thus, the total time is $\sum_{u \in V_G} (deg(u) \cdot |nb_\geq(u)|) \leq \sum_{u \in V_G} (deg(u) \cdot 2\sqrt{m}) = (m \cdot 2\sqrt{m}) = O(m^{1.5})$. Summing up, the total time complexity is $O(m^{1.5})$.

Algorithm 2 uses $O(m+n)$ space to hold the input graph, $O(m)$ space for the hashtable for edge membership test at Step 8, and $O(m)$ space for the edge arrays for the bin sort. Thus, the space complexity is $O(m + n)$. □

Note that the complexity, both time and space, of Algorithm 2 in the worst case is the same as the lower-bound complexity for triangle listing [30].

## 3.3 Limitations of In-Memory Algorithms

Both Algorithm 2 and Algorithm 1 are in-memory algorithms that require $O(m + n)$ memory space. The constant factor is small as we only need extra space for the queue or sorted edge array, the hashtable or for marking whether an edge is deleted, in addition to the space used to hold the input graph. However, for processing a large graph in practice, any small multiplicative factor results in huge amount of extra memory space needed. With the drastically increased volume of graphs/networks in recent years, it is unrealistic to assume that the input graph can always fit in memory.

When the input graph cannot fit in memory, Algorithm 2 and Algorithm 1 reveal that random access to vertices and edges stored on disk is necessary, which can incur prohibitively high I/O cost. The effect of locating an edge to be removed may trigger the removal of other edges and this propagating effect can spread to random locations in the graph. Moreover, the removal of an edge may lead to multiple iterations of support downgrading. These may become serious bottlenecks when the relevant parts of the graph do not reside in memory. Therefore, identifying distinct iterative steps and the relevant subgraphs for such iterations can be a key factor in reducing the disk I/O cost for truss decomposition. How to design I/O efficient algorithms is the main focus of the subsequent sections.

## 4. I/O-EFFICIENT DECOMPOSITION

In Sections 5 and 6, we will present two different algorithms that aim at reducing the I/O cost for truss decomposition in large graphs that cannot fit in memory. We briefly discuss the main objectives of the two algorithms and their differences in this section.

- **Bottom-up approach.** The bottom-up approach starts from the smallest $k$, i.e., $k=2$. The algorithm determines a lower bound on the truss number of the edges in $G$, extracts a candidate subgraph that contains all edges in the $k$-class, $\Phi_k$, computes $\Phi_k$ in memory, removes all edges in $\Phi_k$ from the input graph $G$, and then moves on to compute $\Phi_{k+1}$. This process repeats iteratively until all edges in $G$ are removed.

- **Top-down approach.** The top-down approach starts from largest possible $k$. The algorithm determines an upper bound on the truss number of the edges in $G$, extracts a candidate subgraph that contains all edges in $\Phi_k$, computes $\Phi_k$ in memory, removes all irrelevant edges from $G$, and then moves on to compute $\Phi_{k-1}$.

The bottom-up approach extracts a smaller candidate subgraph than the top-down approach in most cases and is also more effective in removing irrelevant edges for the computation of the subsequent $k$-classes. The top-down approach, though less efficient when all the $k$-classes need to be computed, is particularly suitable for applications that demand only the top $k$-trusses, i.e., the $k$-trusses with the largest values of $k$, which is reasonable since those top $k$-trusses are the more important and core part of a graph or network.

The main idea of both bottom-up and top-down approaches is simple but there are a number of technical challenges: (1) all the steps (e.g., lower/upper bound estimation, candidate subgraph extraction) need to be made I/O-efficient, and it is not straightforward to avoid random access in these steps; (2) many steps in our algorithms apply local computation of global result, ensuring the correctness of the $k$-class globally is a challenge; and (3) pruning is essential for truss decomposition in large graphs but effective and correct pruning is difficult, especially for the top-down approach, due to the tight inter-connections among edges via triangles. We address all these issues in our algorithms.

## 5. BOTTOM-UP TRUSS DECOMPOSITION

In this section, we discuss in details the bottom-up approach for truss decomposition. We first outline the framework of the algorithm, which consists of the following two main stages.

**Lower-bounding:** This stage forms the basis for the later stage of truss decomposition. We compute a nontrivial lower bound on the truss number of each edge. We also remove the 2-class, $\Phi_2$, which can be directly obtained from lower-bounding, to reduce the input size to the subsequent computation.

**Bottom-up truss decomposition:** This stage computes the $k$-classes iteratively bottom-up, i.e., start from the 3-class to the $k_{max}$-class. We use the lower bound to extract a small candidate subgraph, which can be loaded in main memory in most cases to avoid random disk access. After computing each $k$-class, we remove all the edges in the $k$-class, thus reducing the costs for computing the remaining $k$-classes.

### 5.1 Lower Bounding

We first define the concept of *neighborhood subgraph* of a set of vertices, which is frequently used in our algorithms.

*Definition* 4 (NEIGHBORHOOD SUBGRAPH). *Let $U \subseteq V_G$. The* neighborhood subgraph *of $U$, denoted by $NS(U)$, is a subgraph of $G$ which is defined as $NS(U) = (V_{NS(U)}, E_{NS(U)})$, where $V_{NS(U)} = U \cup \{v : v \in nb(u), u \in U\}$ and $E_{NS(U)} = \{(u,v) : (u,v) \in E_G, u \in U\}$.*

**Algorithm 3** *LowerBounding*

**Input**: $G = (V_G, E_G)$
**Output**: $\Phi_2$, and a new graph $G_{new}$, each edge $e$ in $G_{new}$ is associated with a lower-bound on the truss number, $\varphi(e)$

1. $\forall e \in E_G: \varphi(e) \leftarrow 0$;
2. **while**(*not* all edges in $G$ are removed)
3.    partition $V_G$ into $\mathcal{P} = \{P_1, P_2, \ldots, P_p\}$,
     s.t. each $P_i \in \mathcal{P}$ fits in memory;
4.    **for** each $P_i \in \mathcal{P}$ **do**
5.      let $H$ be the neighborhood subgraph $NS(P_i)$ of $P_i$;
6.      compute $\phi(e, H)$ for each edge $e$ in $H$ by Algorithm 2;
7.      $\varphi(e) \leftarrow max\{\varphi(e), \phi(e, H)\}$ for each edge $e$ in $H$;
8.      $\Phi'_2 \leftarrow \{$internal edge $e$ of $H : sup(e) = 0\}$;
9.      output $\Phi'_2$ as part of $\Phi_2$,
       and remove $\Phi'_2$ from $H$ and $G$;
10.     output each remaining internal edge $e$ of $H$, with $\varphi(e)$,
      as part of $G_{new}$, and remove $e$ from $G$;

Intuitively, the neighborhood subgraph of $U$ is a subgraph obtained by adding to $G[U]$ those edges from the vertices in $U$ to those neighboring vertices of $U$ in $V_G \backslash U$, where $G[U]$ is the induced subgraph of $G$ by $U$. We refer to $U$ and $E_{G[U]}$ as the *internal vertices* and *internal edges* of $NS(U)$, respectively, while the remaining vertices and edges in $NS(U)$ are called *external vertices* and *external edges*.

We now outline the algorithm for lower-bounding the truss number of the edges in Algorithm 3. The algorithm iterates over Steps 2-10, computes the lower-bound on the truss numbers for a portion of edges at each iteration, and removes these edges, until all edges in the input graph $G$ are removed.

At each iteration, we partition the vertex set of the *current* graph $G$ (note that $G$ is shrinking at each iteration) into $p \geq 2|G|/M$ parts (Step 3). Chu and Cheng proposed three linear-time partitioning algorithms for triangle listing in large graphs that cannot fit in memory [13]. The first one sequentially partitions $G$, which is fast but does not have a theoretical guarantee on the number of iterations. The second one uses a dominating vertex set of $G$ as seeds to guide the partitioning, which uses $O(n)$ memory space but the number of iterations can be bounded by $O(m/M)$. The third one is a randomized partitioning algorithm that removes the space requirement of the second algorithm and still bounds the number of iterations by $O(m/M)$ with high probability. Since their algorithms partition $G$ into $p$ approximately equal-sized neighborhood subgraphs, where each subgraph fits in main memory, we can apply any of them in our algorithm.

For each of the $p$ neighborhood subgraphs, $H$, we compute the truss number $\phi(e, H)$ for each edge $e$ locally in $H$. Since $H$ fits in main memory, we apply the in-memory truss decomposition algorithm, Algorithm 2. Then, we assign $\phi(e, H)$ as the lower bound $\varphi(e)$ of the truss number of $e$ globally as in $G$. The following lemma establishes the relationship between the local truss number $\phi(e, H)$ in $H$ and the global lower bound $\varphi(e)$ in $G$.

LEMMA 1. *Given a graph $G$ and any neighborhood subgraph $H$ of $G$, we have $\phi(e) \geq \phi(e, H)$.*

PROOF. The proof follows directly from the fact that $H$ is a subgraph of $G$. □

Lemma 1 implies that the maximum $\phi(e, H)$ of any neighborhood subgraph $H$ (at any iteration) can be used as a lower bound of the truss number of an edge $e$ in $G$, as done in Steps 6-7.

In the process of computing $\phi(e, H)$ by Algorithm 2, we can also obtain $sup(e)$ for each internal edge $e = (u, v)$ in $H$. Note that $sup(e)$ computed locally in $H$ is the exact support of $e$ globally in the current $G$, since both $nb(u)$ and $nb(v)$ are in $H$ as $u$ and $v$ are internal vertices. Here in Steps 8-9, however, we only need to determine whether $sup(e) = 0$ for an internal edge $e$ in $H$. If $sup(e) = 0$, then by definition of truss $e$ belongs to the 2-class $\Phi_2$. Thus, we output all internal edges in $H$ with support 0 as the 2-class and also remove them from $G$ to reduce the search space in the subsequent iterations.

Then at Step 10, we output each remaining internal edge $e$ of $H$, with lower bound $\varphi(e)$. All these edges form a new graph, $G_{new}$, which is stored as a list of edges on disk.

At the end of each iteration, all internal edges of each neighborhood subgraph $H$ are removed from $G$. At the next iteration when the shrunk graph $G$ is re-partitioned, some of the external edges in the previous $G$ at the previous iteration now become internal edges, which are processed and finally also removed at the end of the current iteration. This process continues until the shrunk graph $G$ can finally fit in main memory, in which case all edges are internal edges.

## 5.2 Bottom-Up Truss Decomposition

The second stage of the bottom-up approach is the process of bottom-up truss decomposition in the new graph $G_{new}$ obtained by Algorithm 3, where each edge $e$ in $G_{new}$ is associated with a lower-bound on the truss number $\varphi(e)$. We give the algorithm outline in Algorithm 4 and Procedure 5.

Algorithm 4 starts from $k = 3$, extracts a candidate subgraph $H$ for computing $\Phi_k$ (Steps 3-5), compute $\Phi_k$ from $H$ (Step 6), and then moves on to $(k+1)$, until all edges in $G_{new}$ are removed (edges are removed at Step 6).

The candidate subgraph $H$ is extracted from $G_{new}$ as follows. We first scan $G_{new}$ once to obtain the set of candidate vertices $U_k$. Then, we scan $G_{new}$ again to extract all edges that have at least one end vertex in $U_k$, which gives $H = NS(U_k)$. If $U_k$ fits in memory, a single scan of $G_{new}$ suffices.[1] We also need to convert the set of edges extracted from $G_{new}$ into the adjacency list representation. If $H$ fits in memory, the conversion is straightforward.[2]

The correctness of the candidate subgraph for computing $\Phi_k$ will be proved in Theorem 2. We now discuss the main step in Algorithm 4, i.e., Step 6, which computes $\Phi_k$ from the candidate subgraph $H$ by Procedure 5 as follows.

Steps 1-5 of Procedure 5 are similar to Steps 2-11 of Algorithm 2, except that Procedure 5 only computes $\Phi_k$ and focuses on the internal edges of $H$. After outputting an edge $e$ as a $\Phi_k$ edge, we remove $e$ from $H$ to reduce search space. The correctness and completeness of $\Phi_k$ computed will be proved in Theorem 2.

At the end of the procedure, we also remove all edges in $\Phi_k$ from $G_{new}$ to reduce both search space and disk I/O cost for computing the remaining $k$-classes. This step requires reading $G_{new}$ and re-writing the reduced $G_{new}$ back to disk. Checking whether an edge of $G_{new}$ is in $\Phi_k$ can be done efficiently by keeping $\Phi_k$ in a hashtable in memory. If $\Phi_k$ cannot fit in memory, we need $|\Phi_k|/M$ scans of $G_{new}$ to remove all edges in $\Phi_k$ from $G_{new}$.

---
[1]It is very rare that even $U_k$ cannot fit in main memory of an ordinary PC today, for which case $G_{new}$ and hence the input graph $G$ must be extra-ordinarily large. In this case, we need $|U_k|/M$ scans of $G_{new}$ to extract $H$.

[2]Otherwise, we can first write the edge list of $H$ to disk when it is extracted from $G_{new}$. Next, we read in $H$ again and distribute the edges into different buckets on disk, according to the end vertices. Then, we read in each bucket, which fits in memory (otherwise we can use smaller buckets). And finally, we do the conversion for the edges in the bucket in memory and write the converted adjacency list to disk.

**Algorithm 4** *Bottom-Up Truss Decomposition*

**Input**: $G = (V_G, E_G)$
**Output**: the $k$-class, $\Phi_k$, for $2 \leq k \leq k_{max}$

1. call *LowerBounding* (i.e., Algorithm 3);
2. $k \leftarrow 3$;
3. let $U_k = \{v : v \in V_{G_{new}}, \exists e = (u,v) \in E_{G_{new}}, \text{s.t. } \varphi(e) \leq k\}$;
4. let $H$ be the neighborhood subgraph $NS(U_k)$ of $U_k$;
5. scan $G_{new}$ to extract $H$;
6. call *Bottom-Up-Procedure*$(H, k)$;
7. **if**(*not* all edges in $G_{new}$ are removed)
8. $\quad k \leftarrow (k+1)$;
9. $\quad$ **goto** Step 3;

---

**Procedure 5** *Bottom-Up-Procedure*$(H, k)$

1. compute $sup(e)$ for each internal edge $e$ of $H$ in memory;
2. **while**($\exists$ internal edge $e = (u,v)$ of $H$ s.t. $sup(e) \leq k-2$)
3. $\quad$ output $e$ as a $\Phi_k$ edge;
4. $\quad$ **for** each triangle $\triangle_{uvw}$ in $H$ containing $e$ **do**
5. $\quad\quad$ decrease the support of $(u,w)$ and $(v,w)$ by 1;
6. $\quad$ remove $e$ from $H$ and $G_{new}$;

---

In the above discussion, we have assumed that $H$ can fit in memory, which is true in most cases. When the input graph is very large or the available memory size is small, $H$ may not fit in memory. In this case, we need to scan $H$ multiple times to compute $\Phi_k$, in a similar way as in Algorithm 3. We partition $H$ into $p = 2|H|/M$ subgraphs and compute each subgraph in memory. The computation in each subgraph in memory is the same as in Procedure 5. Since we focus on the internal edges of each subgraph only, iteratively processing the above steps in memory finds all internal edges of $H$ that are in $\Phi_k$. The detailed algorithm description is omitted here for simplicity of presentation but is attached in Appendix.

The following example illustrates how bottom-up truss decomposition is processed.

*Example 3.* We illustrate the two stages of the bottom-up algorithm on our running example in Figure 2. In stage 1, Algorithm 3 partitions $G$ into three subgraphs $P_1, P_2, P_3$ as shown in Figure 3. Initially all the edges $e$ of $G$ have a lower bound $\varphi(e) = 2$. Let $\Phi_k(G')$ be the $k$-class of graph $G'$. Given $NS(P_1)$, Algorithm 2 returns $\Phi_2(P_1) = \{(d,l),(g,l)\}$. All the remaining edges in $NS(P_1)$ belong to $\Phi_4(P_1)$, and $\varphi(e)$ for each such edge $e$ is set to 4. In $NS(P_2)$, the local classes are computed as $\Phi_2(P_2) = \{(f,i),(f,j)\}$ and all the other edges in $NS(P_2)$ belong to $\Phi_3(P_2)$; the lower bound $\varphi(e)$ of each edge $e$ in $\Phi_3(P_2)$ is increased from 2 to 3. $P_3$ is processed in the same way. We add the internal edge $(i,k)$ of $NS(P_3)$ to $\Phi_2$ and remove it, and update the lower bounds of the 6 edges in the clique $\{f,h,i,j\}$ to 4. At the end of stage 1, all edges which have not been removed are written to disk as $G_{new}$. In stage 2, we compute the $k$-classes for $k \geq 3$. To compute the 3-class, it is sufficient to load into main memory the subgraph $NS(U_3)$, which is shown in Figure 4(a). Procedure 5 computes the 3-class from this graph, giving $\Phi_3 = \{(d,g),(d,l),(g,l),(g,k),(h,g),(d,k),(e,f),(e,g),(f,g)\}$. We remove $\Phi_3$ from $G_{new}$. We check that $G_{new}$ is not empty at Line 7 of Algorithm 4, so continue with $k = 4$. $NS(U_4)$ is loaded into memory. It is shown in Figure 4(b). $\Phi_4$ is determined in this step to be the 6 edges in the clique $\{f,h,i,j\}$. This is deleted from $G_{new}$. The remaining graph will trigger another call of Procedure 5 at which point the 5-class will be determined. $\square$

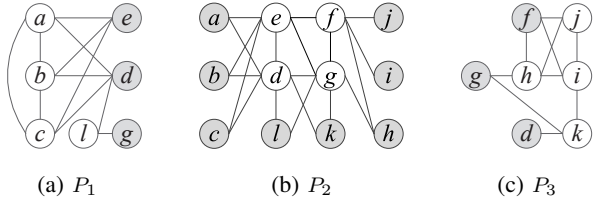

(a) $P_1$  (b) $P_2$  (c) $P_3$

**Figure 3: A partition $\mathcal{P} = \{P_1, P_2, P_3\}$ of $G$ where $P_1 = \{a,b,c,l\}, P_2 = \{d,e,f,g\}, P_3 = \{h,i,j,k\}$**

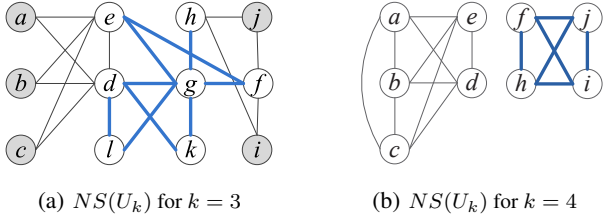

(a) $NS(U_k)$ for $k=3$  (b) $NS(U_k)$ for $k=4$

**Figure 4: Subgraphs of relevant vertices for bottom-up steps**

## 5.3 Correctness and Complexity

We first prove the correctness of Algorithm 4 as follows.

THEOREM 2. *Given a graph $G$, Algorithm 4 correctly computes $\Phi_k$ in $G$, for $2 \leq k \leq k_{max}$.*

PROOF. First, $\Phi_2$ is correctly computed since Algorithm 3 computes the exact support (in $G$) of each internal edge of a subgraph $H$ and we output all internal edges with support 0 as $\Phi_2$ edges. All remaining edges in $G$ must have support at least 1 and thus cannot be in $\Phi_2$. The remaining edges are written to disk as $G_{new}$.

Next, we prove each $\Phi_k$, where $3 \leq k \leq k_{max}$, is computed correctly. We first make an assumption, *Assumption 1*: the candidate subgraph $H$ contains all edges of $\Phi_k$ as internal edges. If Assumption 1 is true, then clearly Procedure 5 (or Procedure 9 in Appendix if $H$ cannot fit in memory) correctly computes $\Phi_k$ since the procedures simply follow the definition of $\Phi_k$. Thus, what remains to be proved is Assumption 1.

Let us make another assumption, *Assumption 2*: $G_{new}$ contains all edges of $\Phi_k$ at Step 3 of Algorithm 4 when $H$ is to be extracted for $\Phi_k$. At Steps 3-5 of Algorithm 4, we extract all edges $e$ with $\varphi(e) \leq k$ from $G_{new}$ as well as their neighboring edges that may form triangles with $e$. From Lemma 1, all edges with $\phi(e) = k$ will be extracted as internal edges. Thus, if Assumption 2 is true, then Assumption 1 must be also true.

Now, we show that Assumption 2 is also true. When $k=3$, Assumption 2 is clearly true since initially $G_{new}$ contains all edges in $\Phi_k$, for $k \geq 3$. Referring to the algorithm, $G_{new}$ is only modified when some $\Phi_k$ is computed and all edges in $\Phi_k$ are removed at Step 6 of Procedure 5 (or Step 12 of Procedure 9 in Appendix). Since all edges in $\Phi_k$ are not used in the computation of $\Phi_j$, for all $j > k$, removing all edges of $\Phi_k$ from $G_{new}$ does not take away any edge in $\Phi_j$. Thus, Assumption 2 is also true and we have established the correctness of Algorithm 4. □

Next, we analyze the complexity of Algorithm 4.

THEOREM 3. *Let $G$ be the input graph. Let $\mathcal{H}$ be the set of candidate subgraphs extracted at Steps 3-5 of Algorithm 4. Let $\mathcal{H}' \subseteq \mathcal{H}$ be the subset of candidate subgraphs that cannot fit in memory. Algorithm 4 computes $\Phi_k$ in $G$, for $2 \leq k \leq k_{max}$, using $O((\frac{m}{M} + k_{max})scan(|G|) + \sum_{H \in \mathcal{H}'} |\triangle_H|)$ I/Os and $O(m^{1.5} + \sum_{H \in \mathcal{H}} |\triangle_H|)$ CPU time.*

PROOF. Algorithm 4 calls Algorithm 3 once, which has the same I/O complexity as that of I/O-efficient triangle listing algorithm [13], i.e., $O(\frac{m}{M} scan(|G|))$ I/Os. If a candidate subgraph $H$ fits in memory, extracting $H$ at Steps 3-5 of Algorithm 4 and processing Procedure 5 require $O(scan(|G_{new}|)) = O(scan(|G|))$ I/Os. For each $H \in \mathcal{H}'$, where $H$ cannot fit in memory, we take a heuristic approach, i.e., Procedure 9, which requires only a few scans of $H$ for real graphs in practice. In the worst case, if the procedure does not terminate after $c$ scans of $H$ for some constant $c$, we can simply take a naive approach by removing the edges with lowest support and the corresponding triangles one by one, which gives the worst case I/O complexity of $O(|\triangle_H|)$.

Algorithm 3 uses $O(|\triangle_G|) = O(m^{1.5})$ CPU time in the worst case, since its complexity is the same as triangle listing [13]. For the computation of each $\Phi_k$, we at most enumerate all triangles in the corresponding $H$, which gives $O(\sum_{H \in \mathcal{H}} |\triangle_H|)$ CPU time. □

In the worst case, $|\triangle_H| = O(|E_H|^{1.5})$. In practice, $|\triangle_H|$ is significantly smaller than $|E_H|^{1.5}$ for real world sparse graphs.

## 6. TOP-DOWN TRUSS DECOMPOSITION

In many applications, one may not want all $k$-trusses; instead, one may be interested in only the top-$t$ $k$-trusses for $k > (k_{max} - t)$. Some applications may even be just interested in the top-1 $k$-truss, i.e., the $k_{max}$-truss, since it represents the very heart of a network/graph. For such applications, applying the bottom-up approach can be wasteful. Therefore, we propose a top-down approach as a solution.

### 6.1 Algorithm Framework

Similar to the bottom-up approach, the framework of the top-down approach also consists of two main parts.

**Upper-bounding:** This part computes a nontrivial upper bound on the truss number of each edge.

**Top-down truss decomposition:** This part computes the top-$t$ $k$-classes iteratively top-down, i.e., from the $k_{max}$-class down to the $(k_{max} - t + 1)$-class. We use the upper bound of the truss number to extract a small candidate subgraph to compute a $k$-class. After obtaining a $k$-class, we remove those edges that do not affect the computation of any $k'$-classes, for $k' < k$, in order to enhance the I/O and CPU performance for computing the $k'$-classes.

Note that the above framework is similar to the top-down core decomposition framework [9]. However, the definition of $k$-truss is based on triangles instead of simple vertex degree as in $k$-core. Thus, the detailed operations in the algorithm are considerably different. In particular, the top-down approach was shown to be more effective than the bottom-up approach for core decomposition [9], but is much less efficient for truss decomposition. This is mainly because the computed edges cannot be effectively removed (unlike the computed vertices in core decomposition) due to the more intricate definition of $k$-truss. Therefore, the top-down approach is only effective for computing the top-$t$ $k$-classes, for some reasonably small $t$.

### 6.2 Upper Bounding

To avoid operating on the entire input graph, the top-down approach first obtains a candidate subgraph for computing the $k$-class

**Procedure 6** *UpperBounding*($G_{new}$)

1. partition $V_{G_{new}}$ into $\mathcal{P} = \{P_1, P_2, \ldots, P_p\}$,
   s.t. each $NS(P_i)$ fits in memory;
2. **for** each $P_i \in \mathcal{P}$ **do**
3.    let $H$ be the neighborhood subgraph $NS(P_i)$ of $P_i$;
4.    **for** each internal edge $e = (u, v)$ of $H$ **do**
5.      let $x_u$ (or $x_v$) be the maximum value of $x$ s.t.
      there are $x$ edges incident to $u$ (or $v$),
      excluding $(u, v)$, with support at least $x$;
6.      $\psi(e) \leftarrow (min\{sup(e), x_u, x_v\} + 2)$;

---

**Algorithm 7** *Top-Down Truss Decomposition*

**Input**: $G = (V_G, E_G)$
**Output**: the top-$t$ $k$-class, $\Phi_k$, for $k_{max} \geq k > (k_{max} - t)$

1. call Algorithm 3, but computing $sup(e)$ instead of $\varphi(e)$;
2. call *UpperBounding*($G_{new}$);
3. $k \leftarrow max\{\psi(e) : e \in E_{G_{new}}\}$;
4. let $U_k = \{v : v \in V_{G_{new}}, \exists e = (u,v) \in E_{G_{new}}$, s.t. $\psi(e) \geq k$
   and $\forall i > k: e \notin \Phi_i\}$;
5. let $H$ be the neighborhood subgraph $NS(U_k)$ of $U_k$;
6. scan $G_{new}$ to extract $H$;
7. call *Top-Down-Procedure*($H, k$);
8. $k \leftarrow (k - 1)$;
9. **repeat** Steps 4-8 until the top-$t$ $k$-classes are computed
   or $G_{new}$ becomes empty;

---

for a given $k$. To extract a candidate subgraph for this purpose, we need to determine an upper bound of the truss number of the edges in the input graph $G$.

We outline the algorithm for computing the upper bound in Procedure 6. The input to Procedure 6 is a graph, $G_{new}$, where each edge $e$ in $G_{new}$ is associated with $sup(e)$ (computed at Step 1 of Algorithm 7). The algorithm partitions $G_{new}$ into $p \geq 2|G_{new}|/M$ neighborhood subgraphs [13], so that each neighborhood subgraph $H$ fits in memory. Then, for each $H$, the upper bound of the truss number of an internal edge $e = (u, v)$ of $H$, denoted by $\psi(e)$, is computed as follows.

Let $w$ be an end vertex of the edge $e = (u, v)$, i.e., $w = u$ or $w = v$. We compute the maximum value of $x_w$ such that there are $x_w$ edges incident to $w$, excluding $e$, with support at least $x_w$. Then, we set $\psi(e) = (min\{sup(e), x_u, x_v\} + 2)$.

*Example 4.* Consider Figure 2. For each edge $e$ in the 5-class, $\psi(e) = 5$. Next consider $(d, g)$. The support $sup((d, g))$ is 3. $x_d = 3$ since there are 3 edges other than $(d, g)$ incident to $d$ with support 3. However $x_g = 2$ since there are only 2 edges other than $(d, g)$ with support at least 2. Hence $\psi((d, g)) = 2 + 2 = 4$.

The following lemma shows that $\psi(e)$ is an upper bound of $\phi(e)$.

LEMMA 2. *Given a graph $G$ and any neighborhood subgraph $H$ of $G$, we have $\phi(e) \leq \psi(e)$, where $e = (u, v)$ is an internal edge of $H$ and $\psi(e)$ is computed in $H$.*

PROOF. First, since $e$ is an internal edge of $H$, all edges incident on $u$ or $v$ are also present in $H$. Suppose to the contrary that $\phi(e) > \psi(e)$. Then, by the definition of $k$-truss, $e$ is contained in more than $(\psi(e) - 2)$ triangles, which also means that there are more than $(\psi(e) - 2)$ edges incident on both $u$ and $v$ with support more than $(\psi(e) - 2)$. However, this implies that $sup(e) > (\psi(e) - 2)$, $x_u > (\psi(e) - 2)$ and $x_v > (\psi(e) - 2)$, which contradicts the fact that $\psi(e) = (min\{sup(e), x_u, x_v\} + 2)$. □

## 6.3 Enumerating Top-$t$ Truss Classes

The second part of the top-down approach mainly concerns with the computation of the top-$t$ $k$-classes, from $k_{max}$ down to $(k_{max} - t + 1)$, for a given $t$, as outlined in Algorithm 7.

The first step in Algorithm 7 calls Algorithm 3. However, since the top-down approach has no need of the lower-bound on the truss number, $\varphi(e)$, of each edge $e$, we do not compute $\varphi(e)$ in Algorithm 3. Instead, we require $sup(e)$ for computing an upper-bound on the truss number, as discussed in Section 6.2. That is, Steps 6-7 of Algorithm 3 are not processed and we replace $\varphi(e)$ by $sup(e)$. However, Algorithm 3 still removes all the edges that belong to $\Phi_2$ to reduce the search space for the later stage of top-down truss decomposition. Note that removing $\Phi_2$ does not affect the support of any other edge that is part of some triangle.

The top-down computation then starts from the largest possible $k$ based on the upper-bounds of the truss numbers, extracts a candidate subgraph $H$ for computing $\Phi_k$ (Steps 4-6), compute $\Phi_k$ from

---

**Procedure 8** *Top-Down-Procedure*($H, k$)

1. compute $sup(e)$ for each internal edge $e$ of $H$ in memory;
2. **while**($\exists$ internal edge $e = (u, v)$ of $H$ s.t. $sup(e) < k - 2$)
3.    **for** each triangle $\triangle_{uvw}$ in $H$ containing $e$ **do**
4.      decrease the support of $(u, w)$ and $(v, w)$ by 1;
5.    remove $e$ from $H$;
6. remove any edge $e \in T_j$ ($j > k$) from $H$ and
   output all remaining internal edges of $H$ as $\Phi_k$;
7. **for** each edge $e = (u, v) \in \Phi_j$, where $e$ in $G_{new}$ and $j \geq k$, **do**
8.    **if**(for every triangle $\triangle_{uvw}$ in $G_{new}$: $\exists i1, i2 \geq k$ s.t.
      $(u, w) \in \Phi_{i1}$ and $(v, w) \in \Phi_{i2}$)
9.      remove $e$ from $G_{new}$;

---

$H$ (Step 7), and then moves on to $(k - 1)$, until all the top-$t$ $k$-classes are computed, or $G_{new}$ becomes empty (in which case all the $k$-classes, for $2 \leq k \leq k_{max}$, are computed).

The candidate subgraph $H$ is extracted from $G_{new}$ in the same way as Steps 3-5 of Algorithm 4, except that $U_k$ is computed based on $\psi(e)$ instead of $\varphi(e)$. However, the first value of $k$ (let it be $k_{1st}$) computed by Step 3 of Algorithm 7 may be much greater than the true $k_{max}$, in which case we will need to repeat Steps 4-8 for $(k_{1st} - k_{max})$ times before the first non-empty $k$-class, i.e., $\Phi_{k_{max}}$, is computed. To avoid this, we may use the smallest $k$ (let it be $k_{init}$) such that the corresponding candidate subgraph $H$ fits in memory. We may then simply apply the in-memory truss decomposition algorithm to compute all $k$-classes, for $k$ in the range from $k_{init}$ to $k_{1st}$. Then, we start the top-down process from $(k_{init} - 1)$.

Step 7 of Algorithm 7 calls Procedure 8 to compute $\Phi_k$ from the candidate subgraph $H$. The algorithm is similar to the bottom-up procedure given in Procedure 5, except for the following differences. First, we iteratively remove all internal edges of $H$ with support less than $(k - 2)$, and then output the remaining internal edges as $\Phi_k$. Second, we remove an edge $e$ from $G_{new}$ only if it is no longer involved in any triangle that contains an edge whose truss number is yet to be computed.

Procedure 8 assumes that $H$ can fit in main memory. We process the case that $H$ cannot fit in main memory in a similar way as we do for the same case in the bottom-up computation, as discussed in Section 5.2. A similar version of Procedure 8 for handling the case that $H$ cannot fit in memory is given in the Appendix.

The following example illustrates how top-down truss decomposition is processed.

*Example 5.* Consider top-down truss decomposition for our running example in Figure 2. After Step 1 of Algorithm 7, $G_{new}$ contains all edges in $G$ except for $(i, k)$. After calling Procedure 6, *UpperBounding*($G_{new}$), we get all the $\psi(e)$ values for edges $e$ in $G_{new}$, and $k$ is set to 5 in Step 4 of Algorithm 7. The induced subgraph of $U_5$ in $G_{new}$ is shown in Figure 5(a). $H = NS(U_5)$ contains also vertices $f, g, l, k$. Next we call Procedure 8 to extract $\Phi_5$, which is determined to be the graph in Figure 5(a). We can remove all of these edges except for $(e, d)$, which is involved in $\triangle_{edg}$ and the truss numbers of $(e, g)$ and $(d, g)$ are not $\geq 5$. Next we set $k = 4$, and repeat Steps 4-8 of Algorithm 7. The induced subgraph of $U_4$ in $G_{new}$ is shown in Figure 5(b). Note that the upper bounds $\psi(e)$ for all edges here are 4 except for (e,d), $\psi((e, d)) = 5$. $H = NS(U_4)$ also contains vertices $l$ and $k$. Next we run Procedure 8 to compute $\Phi_4$, which is the set of edges in the clique $\{f, h, i, j\}$. The 6 edges involved are then removed from $G_{new}$ except for $(f, h)$. Since $t = 2$, the algorithm terminates here.

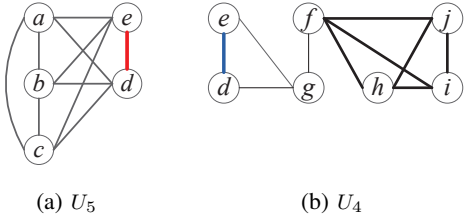

(a) $U_5$          (b) $U_4$

**Figure 5: Top-Down Truss Decomposition for $t = 2$**

## 6.4 Correctness and Complexity

We first prove the correctness of Algorithm 7 as follows.

THEOREM 4. *Given a graph $G$, Algorithm 7 correctly computes the top-t $\Phi_k$ in $G$, for $k_{max} \geq k > (k_{max} - t)$.*

PROOF. In order to prove the correctness of Algorithm 7, we prove that Procedure 8 (or Procedure 10 in the Appendix) correctly computes $\Phi_k$, we need to show that the following statement holds, *Statement 1*: the candidate subgraph $H$ contains all edges of $\Phi_k$ as internal edges and all edges that form triangles with any edge in $\Phi_k$, let us refer to this set of edges as $\Phi_k^\triangle$. If Statement 1 is true, then Procedure 8 (10) returns $\Phi_k$ since these procedures simply operate by following the definition of $\Phi_k$. Statement 1 is true if $G_{new}$ contains $\Phi_k^\triangle$ because $H$ is the neighborhood subgraph of all edges $e$ in $G_{new}$ with unknown truss number where $\varphi(e) \geq k$. From Lemma 2, this set of edges includes all edges of $\Phi_k^\triangle$ if $G_{new}$ contains $\Phi_k^\triangle$ at the time $H$ is to be extracted at Steps 4-6 of Algorithm 7. Next we prove *Statement 2*: $G_{new}$ contains $\Phi_k^\triangle$.

When $k \geq k_{max}$, apparently $G_{new}$ contains $\Phi_k^\triangle$ since no edge has been removed from $G_{new}$ yet. According to Steps 7-9 (or Steps 15-17 of Procedure 10 in Appendix), edges are removed from $G_{new}$ only if they are no longer involved in any triangle that contains an edge whose truss number is yet to be computed. Thus, by induction on $k$ for Statement 2 and the correctness of Algorithm 7, with a base case of $k = k_{max}$, $G_{new}$ always contains $\Phi_k^\triangle$ when $H$ is to be extracted for the enumeration of the $k$-class and the theorem follows. □

Next, we analyze the complexity of Algorithm 7.

THEOREM 5. *Let $G$ be the input graph. Let $\mathcal{H}$ be the set of candidate subgraphs extracted at Steps 4-6 of Algorithm 7. Let $\mathcal{H}' \subseteq \mathcal{H}$ be the subset of candidate subgraphs that cannot fit in memory. Algorithm 7 computes the top-t $\Phi_k$ in $G$, for $k_{max} \geq k > (k_{max} - t)$, using $O((\frac{m}{M} + t)scan(|G|) + \sum_{H \in \mathcal{H}'} |\triangle_H|)$ I/Os and $O(m^{1.5} + \sum_{H \in \mathcal{H}} |\triangle_H|)$ CPU time.*

PROOF. The analysis is similar to that of Theorem 3. □

## 7. EXPERIMENTAL EVALUATION

We compare the performance of our algorithms with the existing in-memory and MapReduce algorithms [15, 16]. All the sequential algorithms were tested on a machine with the Intel Core2 Duo 2.80GHz CPU, 4GB RAM, and the Ubuntu 11.04 operating system. The MapReduce algorithm [16] was ran on an Amazon Elastic MapReduce cluster with 20 nodes, each of which has the computing capacity of a 1.0 GHz 2007 Xeon processor, 1.7GB RAM, and 160GB instance storage; the Hadoop (version 0.20.205) implementation of MapReduce is deployed and the default setting is assumed.

**Datasets.** We use the following nine datasets: *Gnutella Internet peer-to-peer network (P2P)*, *High Energy Physics collaboration network (HEP)*, *Amazon product co-purchasing network (Amazon)*, *Wikipedia Talk network (Wiki)*, *Autonomous systems by Skitter (Skitter)*, *LiveJournal (LJ)*, *Blogs (Blog)*, *Billion Triple Challenge (BTC)*, and *World Wide Web of UK (Web)*. The first six are from the Stanford Network Analysis Project (*snap.stanford.edu*). *P2P* represents hosts as vertices and the connections between the hosts as edges. *HEP* represents each paper as a vertex and each citation between two papers as an edge. *Amazon* represents products as vertices and edges exist between frequently co-purchased products. *Wiki* represents Wikipedia users as vertices and an edge indicates that a user once edited a talk page of another user. *Skitter* describes an internet topology constructed from several sources to about a million destinations. *LJ* is from the free online community LiveJournal (*livejournal.com*), which represents members as vertices and friend hip as edges. *Blog* is from the blogs network and has vertices as blogs and an edge indicates that two blogs appear in the same search result of the top-15 queries published by Technorati (*technorati.com*). *BTC* is an RDF graph cosntructed from the Billion Triple Challenge (*vmlion25.dei.de*). *Web* is from the Yahoo! webspam dataset (*barcelona.research.yahoo.net*) in which webpages are represented as vertices and hyperlinks as edges. Some statistics of the datasets are shown in Table 2. Note that the median degree of most datasets is rather small due to the heavy tail of power-law degree distribution observed in these graphs.

**Table 2: Statistics of datasets ($K = 10^3$, $M = 10^6$, $G = 10^9$): the number of vertices and edges ($|V_G|$ and $|E_G|$), disk storage size (in bytes), maximum and median degree ($d_{max}$ and $d_{med}$), and the largest $k$ for any $k$-truss ($k_{max}$)**

|  | $|V_G|$ | $|E_G|$ | size | $d_{max}$ | $d_{med}$ | $k_{max}$ |
|---|---|---|---|---|---|---|
| P2P | 6.3K | 41.6K | 237K | 97 | 3 | 5 |
| HEP | 9.9K | 52.0K | 317K | 65 | 3 | 32 |
| Amazon | 0.4M | 3.4M | 47.9M | 2752 | 10 | 11 |
| Wiki | 2.4M | 5.0M | 66.5M | 100029 | 1 | 53 |
| Skitter | 1.7M | 11.0M | 149.1M | 35455 | 5 | 68 |
| Blog | 1.0M | 12.8M | 177.2M | 6154 | 2 | 49 |
| LJ | 4.8M | 69M | 809.1M | 20333 | 5 | 362 |
| BTC | 165M | 773M | 10.0G | 1637619 | 1 | 7 |
| Web | 106M | 1092M | 12.2G | 36484 | 2 | 166 |

### 7.1 Performance of In-Memory TD Algorithms

We first assess the efficiency of the improved in-memory algorithm (i.e., Algorithm 2), denoted by **TD-inmem+** (**TD** for Truss Decomposition), compared with the in-memory algorithm [15], denoted by **TD-inmem**.

Table 3 reports the results on the following four datasets, Wiki, Amazon, Skitter, and Blog, which can fit in memory. For the other larger datasets, LJ, BTC, and Web, both algorithms did not

820

complete within reasonable time (longer than a week) due to large amount of memory usage (over 4GB memory).

**Table 3: Running time (wall-clock time in seconds) and peak memory usage (bytes) of TD-inmem+ and TD-inmem**

|  | Wiki | Amazon | Skitter | Blog |
|---|---|---|---|---|
| Time (TD-inmem) | 8856 | 68 | 9204 | 1261 |
| Time (TD-inmem+) | 121 | 31 | 281 | 361 |
| Speedup ratio | 73.2 | 2.2 | 32.8 | 3.5 |
| Mem-usage (TD-inmem) | 966M | 295M | 1.4G | 715M |
| Mem-usage (TD-inmem+) | 846M | 398M | 1.6G | 1.1G |

The results show that TD-inmem+ is apparently much faster than TD-inmem for all datasets. The speedup is from 2.2 times up to 73.2 times faster, with comparable memory consumption. The results suggest that our improved in-memory algorithm not only attains a lower theoretical time complexity, but also significantly improves the efficiency in practice.

## 7.2 Performance of Bottom-Up TD Algorithm

We evaluate the performance of the I/O-efficient bottom-up algorithm, denoted by **TD-bottomup**, compared with Cohen's MapReduce algorithm [16] that also does not require to keep the entire input graph in main memory, denoted by **TD-MR**.

Table 4 reports the running time of TD-bottomup and TD-MR. Note that when the input graph can fit in memory, our I/O-efficient algorithms are simply the in-memory algorithm, TD-inmem+. Thus, we skipped the smaller datasets (which are shown in Table 3) and ran TD-bottomup on the larger networks, LJ, BTC, and Web.

However, we were not able to obtain the result on these large datasets for TD-MR because it is at least more than 3 orders of magnitude slower than our algorithm. In fact, we were only able to obtain the results for TD-MR on the two smallest datasets, HEP and P2P, as shown in Table 4.

**Table 4: Running time (wall-clock time in seconds) of TD-bottomup and TD-MR**

|  | P2P | HEP | LJ | BTC | Web |
|---|---|---|---|---|---|
| Time (TD-bottomup) | <1 | <1 | 664 | 1768 | 6314 |
| Time (TD-MR) | 4200 | 14760 | - | - | - |

The results show that TD-bottomup takes less than 1 second to perform truss decomposition in HEP and P2P, but TD-MR takes 4200 and 14760 seconds on 20 machines. The drastically worse performance of TD-MR is because MapReduce is not a suitable framework for the task of truss decomposition. On the contrary, TD-bottomup uses a single machine and takes less or similar amount of time on datasets with size over 30000 times larger than HEP. The results thus verify that our I/O-efficient algorithm is more feasible for truss decomposition in massive networks.

## 7.3 Performance of Top-Down TD Algorithm

For the top-down algorithm, denoted by **TD-topdown**, we focus on the three large graphs, LJ, BTC, and Web.

We used TD-topdown to compute both the top-20 $k$-classes and all the $k$-classes. Table 5 reports the running time of TD-topdown, where we also show the running time of TD-bottomup as a reference.

The results show that for LJ and Web, the top-down approach has a significant benefit in computing the top-20 results than the bottom-up approach. For the BTC dataset, since $k_{max} < 20$, TD-topdown computes all the $k$-classes and has a comparable speed to that of TD-bottomup. However, TD-topdown is about 6.3 times slower than TD-bottomup for computing all the $k$-classes, and it cannot finish within reasonable time for the largest dataset, i.e., Web. The results thus show that TD-topdown is suitable for computing the top-$t$ results for some small $t$ or for processing datasets that have a small $k_{max}$.

**Table 5: Running time (wall-clock time in seconds) TD-topdown and TD-bottomup**

|  | LJ | BTC | Web |
|---|---|---|---|
| TD-topdown (top-20) | 149 | 1744 | 2354 |
| TD-topdown | 941 | 1744 | - |
| TD-bottomup | 664 | 1768 | 6314 |

## 7.4 K-Truss vs. K-Core

In this experiment, we show that $k$-truss is better than $k$-core as a type of cohesive subgraphs. We compute the $k$-core that has the maximum core number, i.e., the non-empty $k$-core that has the largest value of $k$. We use $c_{max}$ to denote the maximum core number, in order to distinguish it from the maximum truss number (i.e., $k_{max}$), since $c_{max} \neq k_{max}$ in general. We denote the $k_{max}$-truss by $T$ and the $c_{max}$-core by $C$ in this experiment.

**Table 6: Statistics of the $k_{max}$-truss, $T$, and the $c_{max}$-core, $C$, of various datasets (K $= 10^3$): the number of vertices ($V_T/V_C$), the number of edges ($E_T/E_C$), the maximum truss/core number ($k_{max}/c_{max}$), and the clustering coefficient ($CC_T/CC_C$) of $T$ and $C$, respectively**

|  | $V_T/V_C$ | $E_T/E_C$ | $k_{max}/c_{max}$ | $CC_T/CC_C$ |
|---|---|---|---|---|
| Amazon | 5K/33K | 55K/442K | 11/10 | 0.99/0.72 |
| Wiki | 237/700 | 32K/147K | 53/131 | 0.64/0.42 |
| Skitter | 185/222 | 16K/33K | 68/111 | 0.95/0.71 |
| Blog | 49/387 | 2K/54K | 49/86 | 1.00/0.52 |
| LJ | 383/395 | 146K/155K | 362/372 | 1.00/0.99 |
| BTC | 653/1295 | 10K/838K | 7/641 | 0.45/0.00002 |
| Web | 498/862 | 82K/148K | 166/165 | 1.00/0.59 |

Table 6 reports some statistics of $T$ and $C$. The results show that the size of $T$, in terms of both the number of vertices and edges, is significantly smaller than that of $C$, indicating that the "*core*" of a network represented by the $k_{max}$-truss (i.e., $T$) and that by the $c_{max}$-core (i.e., $C$) are radically different.

From another angle, the result that $k_{max}$ is much smaller than $c_{max}$ for most datasets implies that although the $c_{max}$-core is a subgraph in which vertices have a large number of edge-connections to each other, most of these connections do not really form tightly-knit clusters or communities as do in the $k_{max}$-truss. This is verified by the clustering coefficient [33] of $T$ and $C$, i.e., $CC_T$ and $CC_C$ in Table 6. The result clearly indicates that the $k_{max}$-truss and its vertices tend to form clusters much more likely than the $c_{max}$-core.

It is also worth noting that $k_{max}$ and $c_{max}$ are differed by only 1 for Amazon and Web. Since the $k$-truss is a $(k - 1)$-core but not vice versa, e.g., the 11-truss of Amazon is a 10-core or more precisely a subgraph of the 10-core, the result that the 11-truss is so much smaller than the 10-core reveals that out of a large part of a network seemingly to be well-connected, it is possible that only a small portion is truly tightly connected.

The $k$-core can be used as an effective heuristic for maximal clique enumeration [17], since a clique of size $k$ must be in a $(k - 1)$-core, which can be significantly smaller than the original graph. However, our result shows that the $k$-truss can be a much better candidate since its size is in general much smaller than the $k$-core. Note that triangles are fundamental units in a clique and a clique of size $k$ must be in a $k$-truss.

We also know that the size of the maximum clique is bounded by $(c_{max} + 1)$ and $k_{max}$. Thus, our result shows that $k_{max}$ gives a much lower upper bound on the size of the maximum clique. For example, we know that the maximum clique in the Wiki graph has at most 53 vertices as bounded by $k_{max}$, instead of 132 vertices as bounded by $(c_{max} + 1)$.

In conclusion, the results demonstrate that the "*core*" of a network represented by the $k_{max}$-truss (i.e., $T$) is significantly more cohesive or tightly-knit than that by the $c_{max}$-core (i.e., $C$). The results verify that triangle-based connection (as in $k$-truss) is more robust than edge-based connection (as in $k$-core). As cohesive subgraphs are useful for studying important properties (e.g., connectivity, robustness, centrality, etc.) of a network, $k$-truss is a better candidate than $k$-core for network analysis and related tasks. In addition, we also show that $k$-truss can be employed as a more effective heuristic for maximal clique enumeration and maximum clique finding.

## 8. RELATED WORK

The most closely related works to $k$-truss are the cohesive subgraphs [18], such as clique [21, 7], $n$-clique [22], $k$-plex [29], $n$-clan [24], *n-club* [24], various types of quasi-cliques [23, 1], and $k$-core [28], which we have discussed in Section 1.

In addition to the cohesive subgraphs, $k$-truss is also related to dense subgraphs, in particular, the DN-graph [31], which is a connected subgraph in which the lower bound on the number of triangles of edges is locally maximized. Their definition renders the problem NP-hard and their solution is approximate.

The only existing algorithms for truss decomposition were proposed by Cohen [15, 16]. Their first algorithm is an in-memory algorithm [15], which is slow for handling large real-world graphs with power-law distribution. Our in-memory algorithm improves their algorithm by removing the bottleneck for processing vertices with high degree. Their second algorithm is a MapReduce algorithm [16], which is actually not suitable for the task of truss decomposition due to the iterative process that blocks parallelization. We verified in our experiments that this MapReduce algorithm is inefficient and cannot handle large graphs. On the contrary, our I/O-efficient algorithm handles large graphs efficiently.

The framework of our top-down algorithm is similar to that of the algorithm for core decomposition [9], but the detailed design of our algorithms is totally different from theirs and also more complicated than theirs. In particular, the strong triangular connection within the $k$-truss does not allow effective pruning in top-down truss decomposition as in top-down core decomposition [9]. As a solution for finding $k$-truss for all $k$, this paper proposes a more effective bottom-up approach. In direct contrast, the bottom-up approach is not suitable for core decomposition in large graphs [9], since the $k$-cores for small $k$ are generally too large.

Truss decomposition is different from triangle counting [27, 20] since triangle counting is only one step invoked in the iterative process of $k$-truss computation, and any efficient triangle counting algorithm can be applied. In particular, we apply the I/O-efficient algorithms [14, 13] to compute the support of edges. However, ensuring I/O-efficient support counting is only one part of our algorithms and we still have to ensure all the other steps in truss decomposition also I/O-efficient. For example, the bottom-up/top-down steps for iterative truss decomposition.

Other related work includes I/O-efficient maximal clique enumeration [10, 11, 12]. Their algorithms cannot be extended to compute $k$-truss. In fact, both the graph partitioning step and the subgraphs being extracted out of the partition in their algorithms are different from our algorithms.

## 9. CONCLUSIONS

We proposed efficient algorithms for truss decomposition: an in-memory for fast truss decomposition in networks of moderate size, a bottom-up I/O-efficient algorithm for massive networks that are too large to fit in memory, and a top-down algorithm tailored for applications that prefer the top-$t$ $k$-trusses. We verified by experiments on a range of real datasets that our algorithms significantly outperform the existing in-memory algorithm [15] and the MapReduce algorithm [16] on both small and large networks. We also showed that $k$-truss is more suitable than $k$-core as a cohesive subgraph and it reveals tightly-knit clusters of a network.

## 10. ACKNOWLEDGMENTS

The authors would like to specially thank Ada Wai-Chee Fu for her many constructive suggestions and discussions during the entire process of writing this paper, and for her help in polishing the drafts of the paper. The authors also thank the reviewers of this paper for their comments that have helped improve the quality of the paper significantly. This research is supported in part by the A*STAR Thematic Strategic Research Programme Grants (102 158 0034 and 112 172 0013).

# APPENDIX

Procedure 9 is called at Step 6 of Algorithm 4 in Section 5.2, to compute $\Phi_k$ in the candidate subgraph $H$ for the case when $H$ cannot fit in memory. The framework of Procedure 9 is similar to Algorithm 3, while the computation of the $\Phi_k$ edges at Steps 7-11 of Procedure 9 is similar to that at Steps 1-5 of Procedure 5. The detailed explanation is thus omitted here but can be found in Sections 5.1 and 5.2 where Algorithm 3 and Procedure 5 are discussed.

**Procedure 9** *Bottom-Up-Procedure-2*$(H, k)$

1. $sup(e) \leftarrow 0$ for all internal edges of $H$;
2. $H' \leftarrow H$;
3. **while**(*not* all edges in $H'$ are removed)
4.     partition $V_{H'}$ into $\mathcal{P} = \{P_1, P_2, \ldots, P_p\}$, s.t. each $P_i \in \mathcal{P}$ fits in memory;
5.     **for** each $P_i \in \mathcal{P}$ **do**
6.         let $F$ be the neighborhood subgraph $NS(P_i)$ of $P_i$;
7.         compute $sup(e)$ for each internal edge $e$ of $F$;
8.         **while**($\exists$ internal edge $e = (u, v)$ of $F$, s.t. $sup(e) \leq (k-2)$ )
9.             output $e$ as a $\Phi_k$ edge;
10.            **for** each triangle $\triangle_{uvw}$ in $F$ containing $e$ **do**
11.                 decrease the support of $(u, w)$ and $(v, w)$ by 1;
12.            remove $e$ from $F$, $H$, and $G_{new}$;
13.         remove all internal edges of $F$ from $H'$;
14. repeat Steps 2-13 until all remaining internal edges of $H$ have support greater than $(k-2)$;

Procedure 10 is called at Step 7 of Algorithm 7 in Section 6.3, to compute $\Phi_k$ in the candidate subgraph $H$ for the case when $H$ cannot fit in memory. The framework of Procedure 10 is similar to Procedure 9, while the other computation details are the same as Procedure 8.

**Procedure 10** *Top-Down-Procedure-2*$(H, k)$

1. $sup(e) \leftarrow 0$ for all internal edges of $H$;
2. $H' \leftarrow H$;
3. **while**(*not* all edges in $H'$ are removed)
4.     partition $V_{H'}$ into $\mathcal{P} = \{P_1, P_2, \ldots, P_p\}$, s.t. each $P_i \in \mathcal{P}$ fits in memory;
5.     **for** each $P_i \in \mathcal{P}$ **do**
6.         let $F$ be the neighborhood subgraph $NS(P_i)$ of $P_i$;
7.         compute $sup(e)$ for each internal edge $e$ in $F$;
8.         **while**($\exists$ internal edge $e = (u, v)$ of $F$, s.t. $sup(e) < (k-2)$ )
9.             **for** each triangle $\triangle_{uvw}$ in $F$ containing $e$ **do**
10.                 decrease the support of $(u, w)$ and $(v, w)$ by 1;
11.            remove $e$ from $F$ and $H$;
12.         remove all internal edges of $F$ from $H'$;
13. **repeat** Steps 2-12 until all remaining internal edges of $H$ have support at least $(k-2)$;
14. remove any edge $e \in T_j$ $(j > k)$ from $H$ and output all remaining internal edges of $H$ as $\Phi_k$;
15. **for** each edge $e = (u, v) \in \Phi_j$, where $e$ in $G_{new}$ and $j \geq k$, **do**
16.     **if**(for every triangle $\triangle_{uvw}$ in $G_{new}$: $\exists i1, i2 \geq k$ s.t. $(u, w) \in \Phi_{i1}$ and $(v, w) \in \Phi_{i2}$)
17.         remove $e$ from $G_{new}$;